\documentclass[showpacs,twocolumn,aps]{revtex4}
\usepackage{graphicx}

\begin{document}

\title{Relativistic Derivations of de Broglie and Planck-Einstein Equations}

\author{Fabrizio Logiurato}

\affiliation{INO-CNR and LENS, 50125 Firenze, Sesto Fiorentino, Italy \\
Department of Physics -- Trento University, 38123 Povo, Italy}

\date{30nd July 2012}

\begin{abstract}
We show how Special Relativity sets tight constraints on the form of possible relations that may exist between the four-momentum of a particle and the wave four-vector.  More specifically, we demonstrate that there is just one way, according to Special Relativity, to relate the finite energy and momentum of a corpuscle with the characteristics of a plane wave, frequency, wave vector and amplitude: that is by laws of direct proportionality like de Broglie equation ${\bf p}=\hbar {\bf k}$ and Planck-Einstein equation $E=\hbar \omega$.
\end{abstract}

\maketitle

\section{Introduction}

In the autumn of 1924, the French physicist Louis de Broglie submitted to the judgement of Sorbonne University in Paris one of the most famous PhD thesis in the history of physics \cite{deBroglie1}. His results, as he confessed several decades later, gathered the fruits of many solitary meditations on a conundrum of physics which had bothered him for a long time: the dual wave and particle nature of light. His dissertation
brought together various notes mainly published in the Comptes Rendus de the Acad\'emie des Sciences some months before, between 1923 and 1924, \cite{deBroglie1922}-\cite{deBroglie1924}. De Broglie's supervisor was the renowned physicist Paul Langevin \cite{Danin}.
Langevin sent a copy of de Broglie's manuscript to his good friend Einstein. Einstein remained favourably impressed by the work, both for the courageous and innovative ideas contained in it and for the simple mathematics with which these were advanced. Einstein probably also appreciated the important use of Special Relativity in de Broglie's reasoning.

The development of de Broglie's thought, even confining ourselves to the PhD thesis, is rather complex, and not without contradictory points \cite{MACK}. In this introduction we only recall some of its fundamental aspects, referring to the literature for the necessary details 
(see, e.g.: \cite{Dugas}- \cite{Lochak2}).

At  the beginning of the 1920's de Broglie suggests that  the quanta of light could be completely comparable to other known material particles. For instance, they had to share with those the property of having a rest mass different from zero, although very small \cite{deBroglie1922}. Moreover, if the photons had to
be put on the same conceptual framework of the other particles, it was natural to imagine that particles different from the photons could share with the light the strange dual nature of wave and corpuscle.

Then the fundamental hypothesis of the French physicist was to consider true for all the particles, not only for the quanta of light, the Planck-Einstein law: 

\begin{equation} \label{PlanckEinstein0}
E=h\nu   \,,
\end{equation}

\noindent
where $\nu$ is a frequency and $h$ is the Planck constant. But what is the physical origin of the frequency in the Planck-Einstein formula? He will remember some years after, at the conference by him delivered on the occasion of his receiving the Nobel Prize \cite{deBroglieconf}:

\vskip.2cm
{\small When I started to ponder these difficulties two things struck me in the main.
Firstly the light-quantum theory cannot be regarded as satisfactory since it
defines the energy of a light corpuscle by the relation $E= h \nu$ which contains
a frequency $\nu$. Now a purely corpuscular theory does not contain any
element permitting the definition of a frequency. This reason alone renders
it necessary in the case of light to introduce simultaneously the corpuscle
concept and the concept of periodicity.

On the other hand the determination of the stable motions of the electrons
in the atom involves whole numbers, and so far the only phenomena in
which whole numbers were involved in physics were those of interference
and of eigenvibrations. That suggested the idea to me that electrons themselves
could not be represented as simple corpuscles either, but that a periodicity
had also to be assigned to them too.}
\vskip.2cm

De Broglie initially imagines that the source of the frequency is related with some periodic phenomenon inside the particle \cite{deBroglie1923a}. Consider a particle  with velocity $v$ along the $\bf x$ axis in an inertial frame $S$.
Assuming valid the combination of the Planck-Einstein equation for the photons (\ref{PlanckEinstein0}) and the 
relativistic energy  of the particle,

\begin{equation} \label{EinsteinE}
E=\frac{m_0 c^2}{\sqrt{1- \beta^2}}   \,, \qquad \beta= \frac{v}{c} <1 \,,
\end{equation}

\noindent
he writes that in the rest frame $S_0$ of the particle there must be:

\begin{equation} \label{DB2}
h\nu_{0}  = m_{0} c^2\,, 
\end{equation}

\noindent
where $\nu_{0}$ is the frequency of the supposed inner vibration. However, de Broglie is immediately forced to 
face a problem. Because of the relativistic time dilation, an observer, for whom the particle is moving with velocity $v$, ascribes to the inner vibration a lower frequency $\nu_i$:

\begin{equation} \label{Omfi}
\nu_i  = \nu_{0} \cdot \sqrt{1- \beta^2}   \,. 
\end{equation}

\noindent
However, comparing (\ref{PlanckEinstein0}) with (\ref{EinsteinE}):

\begin{equation} \label{PlanckEinstein0EinsteinE}
h\nu = \frac{m_0 c^2}{\sqrt{1- \beta^2}}   \,, 
\end{equation}

\noindent
and considering the condition (\ref{DB2}), we easily obtain the relativistic formula of transformation between frequencies

\begin{equation} \label{Omfo}
\nu = \nu_{0} \cdot \frac{1}{\sqrt{1- \beta^2}}   \,, 
\end{equation}

\noindent
which  is typically linked with a wave phenomenon: $\nu$ is the frequency of a wave in $S$, while $\nu_0$ is now the frequency of such a wave in the frame $S_0$.

In order to resolve this difficulty, de Broglie assumes the exis\-tence of a  \lq fictitious\rq \, wave associated with the par\-ti\-cle \cite{deBroglie1923a}, with frequency $\nu$ and phase velocity

\begin{equation} \label{vfase}
v_p=\frac{c}{\beta} >c  \,, 
\end{equation}

\noindent
that is

\begin{equation} \label{vevfase}
v \cdot v_p=c^2 \,. 
\end{equation}

\noindent
(The wave is fictitious  because, according to the French physicist, being its speed greater than the speed of light, it cannot transport energy).
In order to justify the assumption in equation (\ref{vfase}), he shows that if the periodic inner phenomenon  and the external wave with phase velocity  (\ref{vfase}) are in phase at a given time, they will be always in phase; i.e. the particle moves within the wave  maintaining its inner vibration in phase with the wave. De Broglie called that  \lq law of the harmony of phases\rq.

His result, according to de Broglie, suggests that \lq any moving body could be accompanied by a wave, and it is impossible to disjoin the motion of the body from the wave propagation\rq \, \cite{deBroglie1924}.
Therefore, he assumes that $\nu$ is the frequency of a plane wave which accompanies the particle, and that  this frequency  is the same which is in the Planck-Einstein equation. 

Let us recall the relativistic momentum of the particle:

\begin{equation} \label{impulsor}
p= \frac{m_0 v}{\sqrt{1- \beta^2}}   \,. 
\end{equation}

\noindent
By comparing (\ref{impulsor}) with (\ref{PlanckEinstein0EinsteinE}) and assuming that this holds in any frame, we may write the momentum  in terms of the wave frequency:

\begin{equation} \label{impulsofreq}
p= \frac{h \nu}{c^2} v  \,. 
\end{equation}

\noindent
But from (\ref{vevfase}) we know that $v=c^2/v_p$, and remembering that for a monochromatic wave the wavelength is $\lambda=v_p/ \nu$, from  (\ref{impulsofreq})  we have the formula that  made de Broglie famous:

\begin{equation} \label{DB1}
p=h/\lambda\,.
\end{equation}

\noindent
Equation (\ref{DB1}) connects the module of the momentum $p$ of a particle with the wavelength $\lambda$ of the associated plane wave through the Planck constant $h$. (For  historical precision, we have to say that de Broglie expressly writes (\ref{DB1})  only in the last chapter of his Ph.D. thesis, in the form $\lambda=h/p$. 
In all his preceding works, the treatment of the Bohr atom included, he always reasons in terms of frequency).

In his second memoir on the Comptes Rendus \cite{deBroglie1923b} de Broglie suggests an experimental verification of his ideas. The wave is conceived as a kind of field that guides the behavior of the particle.
The new dynamics of particles, which he plans to develop, would be to the old classic dynamics as the wave optics is to the geometric optics. According to the author: 

\vskip.2cm
{\small The new principle of dynamics would explain the diffraction of light atoms ({\sl the photons}), no matter how small their number. Moreover, any body in certain cases would be subject to diffraction. A stream of electrons through an opening sufficiently small would show some phenomena of diffraction. It is in this direction that we must look for experimental confirmation of our ideas.}
\vskip.2cm

\noindent
De Broglie's prediction on the wave nature of the electron, the reason for his Nobel Prize, will be confirmed a few years later in the experiments of Davisson, Germer \cite{Davisson}  and Thomson \cite{Thomson} on the diffraction of electrons by crystals \cite{LogiuratoPT2008}.

Today just a few introductory textbooks to quantum theory describe the original way of de Broglie's thinking. Some books contain simplified versions \cite{Wichmann}, \cite{FrenchTaylor}; most of them simply cite 
equation (\ref{DB1}) as a postulate, and its application in the deduction of the energy quantization in Bohr's atom model \cite{Gasiorowicz}, \cite{Cassidy}. 
Perhaps, from the present perspective, many of de Broglie's initial suppositions appear strange (but see \cite{Catillon}). However, rejecting entirely the reasoning of the French physicist, and ignoring completely the history of his formula,  means also missing the relativistic argument, which underlines from the beginning how quantum mechanics is related to  Special Relativity (without having to wait for the Dirac equation, with his description of  spin and 
prediction of  antimatter).
This is a pity, as the power to unify different descriptions of the phenomena is one of the more interesting sides of the physics.

In the next section we report an alternative deduction of the de Broglie relation obtained directly from Lorentz transformations and  Planck-Einstein equation. In Section 3, following de Broglie, Ashby, Miller \cite{AshbyMiller} and Reed \cite{Reed}, we show how Special Relativity puts constraints on the possible formulas that may connect energy and  momentum of a particle  with the wavelength, the frequency and the wave amplitude.

We demonstrate in general that equations like de Broglie's and  Planck-Einstein's are both the only relations allowed by Special Relativity, once we assume the existence of a dependency between the four-momentum of a particle and the wave four-vector of  a  plane wave and that the particle cannot be separated  by its wave.


\section{De Broglie relation from  Special Relativity and  Planck-Einstein relation}

For particles without rest mass ($m_0=0$) such as photons, deducing the de Broglie relation from the Planck-Einstein equation  is straightforward. In fact, following Einstein \cite {Einstein1916}, it is enough to consider the relativistic equation between momentum and energy:

\begin{equation} \label{pE}
p= E/c           \,. 
\end{equation}

\noindent
Putting $E= h \nu$  in ($\ref{pE}$) and recalling that $\lambda= c / \nu$, we quickly obtain ($\ref{DB1}$).

An elegant way to derive the de Broglie relation, for any massive or massless particle,  can be achieved using directly Lorentz transformations \cite{deBroglieM}. We put forward the following assumptions:

\vskip.3cm

\noindent
{\bf Postulate 1.a}: Each particle is associated with a wave phenomenon. 

\vskip.2cm

\noindent
{\bf Postulate 2.a}: In every inertial frame the relation  $E=\hbar \omega$ holds, where $E$ is the energy
of the particle, $\omega= 2 \pi \nu$ is the frequency of the associated wave in that frame and $\hbar=h/2\pi$ is a
relativistically invariant constant.

\vskip.3cm

\noindent
We want to show that:

\vskip.3cm
\noindent
{\bf Theorem a}: According to Special Relativity and the postulates 1.a-2.a, between the momentum and the wave vector  there is necessarily the relation ${\bf p}=\hbar { \bf k}$.  

\vskip.4cm

\noindent
Let $S$ and $S'$  be two inertial frames in relative motion. According to  Postulate 2.a, we assume that in both frames the Planck-Einstein relation applies to any particle:

\begin{equation} \label{PLSS}
E=\hbar \omega \,,  \qquad  E'=\hbar {\omega}' \,.
\end{equation}

\noindent
We introduce, for the particle and the wave, the four-momentum $p^\mu$ and the wave four-vector $k^\mu$, respectively:

\begin{equation} \label{pk}
p^\mu=(E/c, {\bf p}) \,, \qquad   \qquad   \qquad  k^\mu=(\omega /c, {\bf k})  \,,
\end{equation}

\noindent
where the wave vector ${\bf k}$ has modulus $k=2 \pi / \lambda$. We assume, for simplicity, that the frames $S$ and $S'$ have parallel axes to each other and that at time $t=t'=0$  the origins $O$ and $O'$ of the spatial coordinates coincide. Moreover, we suppose  $S'$ to move with respect to $S$ with speed $V<c$ along the direction of the
${\bf x}$ axis, and  the direction of the wave propagation to be along such axis (Fig. 1).

\begin{figure}
\includegraphics[width=8cm]{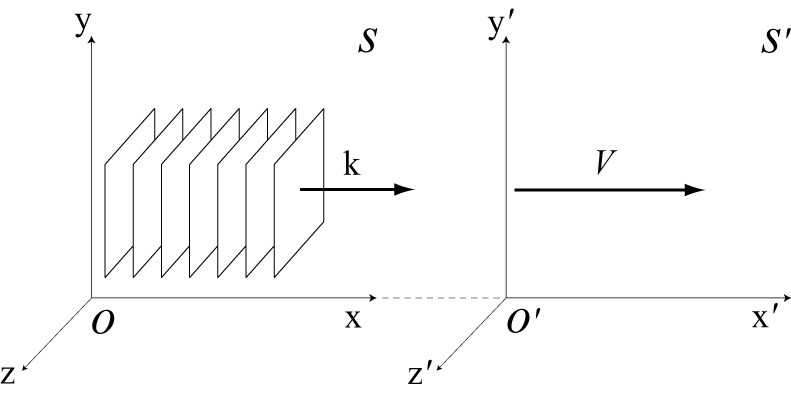}
\caption{
Given a plane wave with wave vector ${\bf k}$, we choose our inertial frame $S$ with the ${\bf x}$-axis coinciding with the direction of the wave vector, so that ${\bf k}=(k_x,0,0)$. The figure schematically shows some fronts of the plane wave. $S'$ is the inertial frame travelling with velocity $V$ in comparison with $S$.} 
\end{figure}


The Lorentz transformations for the four-momentum and the wave four-vector are

\begin{equation} \label{TLQPQK}
\left\{
\begin{array}{rl}
E'/c &= \gamma ( E/c - \beta p_{x}) \\
p'_{x} &= \gamma (p_{x} - \beta E/c) \\
p'_{y} &= p_{y}  \\
p'_{z} &= p_{z}                                               
\end{array}
\right. \qquad 
\left\{
\begin{array}{rl}
\omega'/c &= \gamma (\omega / c  - \beta k_{x})  \\
k'_{x} &= \gamma ( k_{x} - \beta \omega / c)  \\
k'_{y} &= k_{y} \\
k'_{z} &= k_{z} \,,
\end{array}
\right.
\end{equation}

\noindent
where

\begin{equation} \label{defbg}
\beta =  \frac{V}{c} \,, \qquad  \gamma = \sqrt{ 1- \frac{V^2}{c^2}}
\,,
\end{equation}

\noindent
we have

\begin{equation} \label{TLEo}
\left\{
\begin{array}{rl}
E'/c &= \gamma ( E/c - \beta p_{x}) \\
{\omega}'/c &= \gamma (\omega / c  - \beta k_{x})  \,.
\end{array}
\right.
\end{equation}

\noindent
By multiplying the second of ($\ref{TLEo}$)  for the Planck constant $\hbar$:

\begin{equation} \label{TLEp}
\left\{
\begin{array}{rl}
E'/c &= \gamma ( E/c - \beta p_{x}) \\
\hbar {\omega}'/c &= \gamma ( \hbar \omega / c  - \beta \hbar  k_{x})  \,.
\end{array}
\right.
\end{equation}

\noindent
Subtracting side by side the two equations $(\ref{TLEp})$:

\begin{equation} \label{TL3}
E'/c -   \hbar {\omega}'/c = \gamma (E/c - \hbar \omega / c  - \beta p_{x} + \beta \hbar  k_{x})  \,,
\end{equation} 

\noindent
from which, because of ($\ref{PLSS}$) assumed at the beginning, we get

\begin{equation} \label{TL4}
\gamma \beta (p_{x} - \hbar  k_{x})=0  \,.
\end{equation} 

\noindent
If we exclude the trivial condition in which the relative velocity $V$ of the frames is zero (in such case the 
factor $\gamma \beta $ is also zero and the two systems $S$ and $S'$ coincide), equation ($\ref{TL4}$)  is only satisfied with

\begin{equation} \label{TL5}
p_{x} = \hbar k_{x}  \,,
\end{equation} 

\noindent
equivalent to the de Broglie equation for the $x$ component of the momentum.

We point out as the   just given demonstration, with  Postulate 2.a, holds for particles of any mass, while de Broglie's demonstration, starting from the relation (\ref{impulsor}), only holds for particles with nonzero rest mass.

Equation (\ref{TL5})   can be easily generalized to the other components of momentum and wave vector. 
In fact, 
consider a general orientation of the wave vector ${\bf  k}$ with respect to  the $S$ frame $(k_y\neq0, k_z\neq0 )$. 
We may use inertial  frames $S'_y$ and $S'_z$  that  travel with  velocities along ${\bf  y}$ and ${\bf  z}$ with respects to $S$.
Applying the   correspondent relations  of (\ref{TLEo}) for  $S'_y$ and $S'_z$ with the
pairs of components $(E/c, p_y)$, $(\omega/c, k_y)$ and $(E/c, p_z)$, $(\omega/c, k_z)$  
we have at once:

\begin{equation} \label{TL5g}
p_{y} = \hbar k_{y}  \,, \qquad  p_{z} = \hbar k_{z} \,.
\end{equation}

\noindent
The presented deduction can be applied also to other pairs of four-vectors 
in order to show that, if a proportionality relation holds between two components of two four-vectors, then the proportionality must also hold for the other components. That is, given two generic four-vectors $(a_0, a_x)$ and  $(b_0, b_x)$, if it is $a_0=C b_0$, necessarily from Lorentz transformations it has to be $a_x=C b_x$, and vice versa.

Let us consider, as an example, the four-momentum $p^\mu=(E/c, {p_x})$ and the four-velocity $u^\mu=(u_0, {u_x})$,
where $u_0=\gamma c$ and $u_x=\gamma v_x$. Assuming a law of proportionality between ${p_x}$ and ${u_x}$ through an invariant $m_0$ with the dimension of a mass,
$p_x=m_0 u_x$, and arguing in a similar way as reported just above, we can show that $E/c=m_0 u_0$, equivalent to the famous Einstein formula $E= m c^2$, with $m=\gamma m_0$.


\section{De Broglie and Planck-Einstein relations together from  Special Relativity}
\label{PlanckEinstein}

As already remarked by Einstein in one of his fundamental works of 1905, 
the energy of an electromagnetic  radiation contained in a closed surface, and the frequency of the same radiation, change under Lorentz transformations in the same way \cite{Einstein1905}.

As pointed out by Ashby and Miller \cite{AshbyMiller}, the energy of a massless particle and the frequency of an electromagnetic wave
change under transformations in a similar way. 
Let's consider, in fact, once again ($\ref{TLEo}$). By placing in the first $p_x=E/c$ and in the second $k_x=\omega/c$, we have

\begin{equation} \label{TLEofr}
\left\{
\begin{array}{rl}
E'/c &= \gamma ( E/c - \beta E/c) \,, \\
{\omega}'/c &= \gamma (\omega / c  - \beta \omega /c)  \,.
\end{array}
\right.
\end{equation}

\noindent
We get from the previous relations and recalling the definitions ($\ref{defbg}$) of the factors $\beta$ and $\gamma$:

\begin{equation} \label{TLEf}
E'= E \cdot \sqrt{\frac{1-\beta}{1+\beta}} \,,  \qquad  \qquad    
{\omega}'=\omega  \cdot \sqrt{\frac{1-\beta}{1+\beta}} \,.
\end{equation}

\noindent
According to Ashby and Miller  the simultaneous validity of ($\ref{TLEf}$) gives strong constraints on the
possible forms that a relation between the energy of a photon and the frequency of a wave may have.
In fact, these authors assume, ab absurdo, $E= C {\omega}^n$,
where $n$ is in general different from 1 and $C$ is an invariant constant. 
They show that the Planck-Einstein relation, which is obtained with $n=1$,  is the only kind of dependence between energy and frequency which is relativistically invariant.

It would seem, from the developed reasoning in section 2 and from the Ashby and Miller result, that, at least for the photons, both  Planck-Einstein and de Broglie equations may follow from Special Relativity.

Then we wonder if a general constraint exists, which is valid for particles with any mass, for which the condition of relativistic invariance imposes the form of both Planck-Einstein and de Broglie relations.

As a matter of fact, in a not well known work, Reed (1991) has shown  that, if we identify the velocity of a massive particle with the group velocity of  a wave packet, between the four-momentum of the particle and the mean of the wave four-vector there is a relation of proportionality \cite{Reed}. So Reed has deduced together two relations
which are similar to de Broglie's and  Planck-Einstein's.  

However, his demonstration suffers from some defects. In particular, the fact that in quantum mechanics a wave packet is associated with a statistical ensemble of particles which have a dispersion of the values of the momentum; while a particle with a well definite momentum does not correspond to a wave packet, but to a monochromatic wave function \cite{phasev}.

Here, we intend to show that even between the determinate four-momentum of a particle and the wave four-vector of a monochromatic wave there must be a condition of direct proportionality. Our initial hypotheses will be more general in comparison with Reed's. Moreover, differently from Reed, we shall consider particles with and without inertial mass.

We assume from experience that every particle is in relation with a wave \cite{Feynman}. We  make no assumptions about the specific nature of this wave, (still a controversial issue after almost a century \cite{Laloe}) or on the type of differential equation it may have: the d'Alembert wave equation, or Schr\"odinger's, or Klein-Gordon's, or Dirac's, or more. We suppose only that a perturbation of an unclear kind, describable as a plane wave with definite wave number and frequency, is associated with a corpuscole and with the finite energy and the finite momentum of it. 

We recall that in classical mechanics a plane wave possesses infinite total momentum and infinite total energy, and then we can only define for it a flux and a density of momentum, or a flux and a density of energy \cite{Elmore}, so our assumption, that we may call \lq quantistic\rq \, is in contrast with  classical mechanics. 

We consider as previously, for sake of simplicity, a plane wave of angular frequency $\omega$ travelling with wave vector  ${\bf k}$ along the ${\bf x}$ positive direction of an inertial frame $S$. All our results will be generalizable for ${\bf k}$ with generic direction with respect to $S$ following the way sets out at the end of  section 2.

In $S$, $k_y=k_z=0$ and because of Lorentz transformation ($\ref{TLQPQK}$) of the wave four-vector, the $y$ and $z$ components of $\bf{ k}$ will be always zero also in any other inertial reference $S'$.
Therefore we will look for the expressions of $E$ and $p_x$  only as functions  of $ \omega$  and $k_x$. In fact, since we assume that no force acts on our system, the energy and the momentum depend neither from the spatial coordinate $\bf {x}$ nor from the time $t$. Let us denote with $A$ a generic amplitude of the wave (without defining whether it is a scalar, a four-vector or other).
We assume the following postulates:

\vskip.3cm

\noindent
{\bf Postulate 1.b}: Each particle  is associated with a wave and it is impossible to disjoin the motion of the particle from its wave. 

\vskip.2cm

\noindent
{\bf Postulate 2.b}:
The finite energy and momentum of a free particle  are associated with the characteristics of a monochromatic plane wave, amplitude, frequency and wave vector:

\begin{equation} \label{inv1b} 
E=E (A, \omega, k_x) \,, \qquad    p_x=p_x (A, \omega, k_x)   \,.
\end{equation}

\noindent
We intend to show:

\vskip.2cm
\noindent
{\bf Theorem b}: 
The only functions (\ref{inv1b}) allowed by  Special Relativity and by  Postulates 1.b-2.b, are the relations of proportionality $E=C \omega$ and $p_x=C k_x$, where $C$ is a relativistic invariant.  
\vskip.3cm

\vskip.2cm

\noindent
For the development of Theorem b we need the following Lemma:

\vskip.2cm

\noindent
{\bf Lemma}: In accordance with the adopted postulates, if in a frame $S_0$ the momentum of the particle is zero, $p_x=0$, then in that same frame the wave vector must be zero as well, $k_x=0$, and vice versa.  

\vskip.2cm

\noindent
According to our postulates, if the particle is somehow connected to its wave, the momentum has to flow in the  direction of the wave propagation (in conformity with what happens to waves in classical mechanics).

In fact, consider an emitted particle by a source placed at the point $A$ and absorbed by a detector at the point $B$: in such a situation there is a propagation and a transfer of momentum between the source and the detector.
Now imagine the ideal monochromatic wave as the limit case of a wave train with a broad but finite length, by letting its length to grow to infinity, and consequently increasing the distance between points $A$ and $B$. 
If the momentum is linked with the particle, the particle  with the wave, and the wave train must be emitted with the particle, then
the detected momentum must propagate in the  direction of propagation of the wave train, from $A$ to $B$: if $k_x>0$ there is even $p_x>0$, if $k_x<0$ we have $p_x<0$, and vice versa:

\begin{equation} \label{onetoone}
p_x>0   \Longleftrightarrow k_{x}>0 \,, \qquad
p_x<0   \Longleftrightarrow k_x<0  \,.                                          
\end{equation}

\noindent
We require that such a condition holds in every inertial system $S$.
Let $S_0$ be the frame in which the momentum of the particle is zero ($p_{0x}=0$).
Lorentz transformations between the two inertial frames $S$ and $S_0$ for the components of the momentum and the wave vector are:

\begin{equation} \label{TLQPQKBOs0}
p_{x} = \gamma \cdot \frac{v_x}{c} \frac{E_0}{c}\,, \quad
k_{x} = \gamma \cdot ( k_{0x} + \frac{v_x}{c} \frac{\omega_0}{c}) \,,
\end{equation}

\noindent
where now we have represented in explicit form the dependence of $\beta$ on the $v_x$ velocity  of $S_0$, coincident with the rest frame of the particle (hence $v_x$ is also the velocity of the particle with respect to $S$).
The sign of $p_{x}$ only depends on the sign of $v_x$ ($\gamma=\gamma(v_x)>0$). Therefore the condition (\ref{onetoone}) is 
satisfied if, and only if, in $S_0$ is $k_{0x}=0$. It is easy to understand from (\ref{onetoone})  
and Lorentz transformations that, in general, there must be:

\begin{equation} \label{pks0}
p_{0x}=0  \Longleftrightarrow k_{0x}=0 \,.
\end{equation}

\noindent
So, assuming that the momentum (i.e., the velocity of the particle) is always in the direction of the wave propagation, it is equivalent to assuming the existence of a frame in which momentum and wave vector are both zero: the frame where the classical particle is at rest is also the frame where the wave is stationary, (a special kind of stationary wave, with wavelength infinite).

\vskip.2cm

Let us look for the velocity of the frame where the wave is stationary  in function of $\omega$ and $k_x$. Putting 
$k_{0x}=0$ into the inverse of the second transformation in (\ref{TLQPQKBOs0}),

\begin{equation} \label{TLkoin}
k_{0x} = \gamma \cdot (k_{x} - \frac{v_x}{c} \frac{\omega}{c}) \,,
\end{equation}

\noindent
we get immediately:

\begin{equation} \label{TLkoinb}
k_{x} = \frac{v_x}{c} \frac{\omega}{c} \,.
\end{equation}

\noindent
Remembering that the phase velocity  is defined as $v_p=\omega/k_x$, from (\ref{TLkoinb}) we have

\begin{equation} \label{vxvp}
v_x = c^2/v_p \,.
\end{equation}

\noindent
Equation (\ref{vxvp}) is exactly equivalent to (\ref{vevfase}), postulated by de Broglie in order to obtain the harmony of phases between the periodic inner phenomenon and the external wave. From equation (\ref{vxvp}), being the velocity of the  particle $v_x \leq c$, we see that the phase velocity of the associated wave has to be $v_p\geq c$, 
(where the sign of equality applies only to massless particles.).

\vskip.6cm
Coming back to Theorem b, we consider the two invariants square moduli of the four-momentum and the wave four-vector:

\begin{equation} \label{invmcom}
E^2/c^2-p_x^2=m_0^2 c^2  \,, \qquad   \omega^2/c^2-k_x^2=\omega_0^2 /c^2 \,,
\end{equation}

\noindent
where $m_0 c^2$ and $\omega_0$ are, respectively, the energy $E_0$ and the frequency in the frames where the momentum and the wave vector are zero. We study the four different possible cases:

\vskip.8cm

\noindent
{\it Case 1: ${m_0\neq0}$\,,  \qquad ${\omega_0\neq0}$.}
\vskip.2cm

\noindent
We have, from Lorentz transformations between $S$  and $S_0$ for the four-vectors $p^\mu$ and  $k^\mu$:

\begin{equation} \label{TLQPQKB}
\left\{
\begin{array}{rl}
E/c &= \gamma ( E_0/c + \beta p_{0x}) \\
\omega/c &= \gamma (\omega_0 / c  + \beta k_{0x})                                              
\end{array}
\right. \qquad 
\left\{
\begin{array}{rl}
p_{x} &= \gamma (p_{0x} + \beta E_0/c)  \\
k_{x} &= \gamma (k_{0x} + \beta \omega_0 / c) \,.
\end{array}
\right.
\end{equation}

\noindent
Now, according to the Lemma, we assume that in the frame $S_0$ is $p_{0x}=0$ and  $k_{0x}=0$:

\begin{equation} \label{TLQPQKBB0}
\left\{
\begin{array}{rl}
E &= \gamma  E_0 \\
\omega&= \gamma \omega_0                                             
\end{array}
\right. \qquad 
\left\{
\begin{array}{rl}
p_{x} &= \gamma \beta E_0/c  \\
k_{x} &= \gamma \beta  \omega_0 / c \,.
\end{array}
\right.
\end{equation}

\noindent
we get, from (\ref{TLQPQKBB0}), dividing side by side the two equations:

\begin{equation} \label{rapeoo}
\frac{E}{\omega}= \frac{E_0}{\omega_0} \,, \qquad  \qquad  \frac{p_x}{k_x}= \frac{E_0}{\omega_0} \,.
\end{equation}

\noindent
So, being the inertial
frame $S$ arbitrary, from the first of (\ref{rapeoo}) we deduce
that the ratio $E/ \omega$  has to be an invariant. Then we can introduce the invariant $C$:

\begin{equation} \label{invCoo}
\frac{E_0}{\omega_0} \equiv C \,,
\end{equation}

\noindent 
where $C$ is for the initial hypotheses finite, positive and constant with respect to the space-time coordinates $(t, x)$.

\noindent
By confronting  (\ref{invCoo}) with (\ref{rapeoo}), we directly have:
 
\begin{equation} \label{relPDBL4a}
E= C \omega  \,, \qquad  \qquad   p_{x} =C k_{x} \,.
\end{equation}

\noindent
Identifying $C$ with the Planck constant, we can recognize in (\ref{relPDBL4a}), respectively, the Planck-Einstein and the de Broglie relations. 

From the definition of $C$ in  (\ref{invCoo}), we see that 

\begin{equation} \label{invC000}
C=\frac{m_0 c^2}{\omega_0} \,.
\end{equation}

\noindent
Therefore $C$ does not explicitly depend on the amplitude, but may depend on the inertial mass of the particle and the invariant $\omega_0$, the frequency  in the frame where the wave is statio\-nary. In general:

\begin{equation} \label{invC0001}
C=C(m_0, \omega_0) \,.
\end{equation}

\noindent
However, if we require, according the Postulate 2.b and the first of equations (\ref{rapeoo}), that in the limit $m_0\rightarrow 0$ the invariant $C$ is finite and different from zero,
from (\ref{invC000}) it must necessarily be also $\omega_0\rightarrow 0$ (we will analyze the case  $m_0=0$ and $\omega_0=0$  below). This claim implies the existence of a dependence between $\omega_0$ and  $m_0$, $\omega_0 = \omega_0 (m_0)$. So, in this circumstance, $C$ can only depend on  the  mass, $C=C(m_0)$, and
the invariant is constant for that particle. That is equivalent to assuming the inertial mass is the only invariant  which plays a role in the problem.

Finally, if we demand that $C$ is independent of the mass, such as the Planck constant seems to be experimentally \cite{Cronin}, we deduce easily from (\ref{invC000}) that must be $m_0=C'\omega_0$, where $C'=C/c^2$ is an identical constant for every particle. Then the inertial mass is proportional to the frequency of a periodic phenomenon, as initially supposed by de Broglie.

\vskip.8cm

\noindent
{\it Case 2: ${m_0\neq0}$\,,  \qquad ${\omega_0=0}$.}
\vskip.2cm

\noindent
Now, in $S_0$, we have $k_{0x}=0$ and $ \omega_{0}=0$. So, from Lorentz transformations (\ref{TLQPQKB}), if $\gamma$  was finite in every generic frame $S$ there would be $k_x=0$ and $\omega_x=0$, a situation physically impossible for a wave.

A way out is represented by allowing the factor $ \gamma$ to be infinite. But this happens only for $v_x=c$, and in such a situation the frame $S_0$ and the particle travel at the speed of light. In order to avoid the infinite energy of the particle we should assume $m_0=0$, in contradiction with the conditions of  \textit{Case  2}.

\vskip.8cm

\noindent
{\it Case 3: ${m_0=0}$\,,  \qquad ${\omega_0\neq0}$.}
\vskip.2cm

\noindent
In this case, since $E_0=m_0 c^2$, in $S_0$ we have $E_{0}=0$ and $p_{0x}=0$. Then, from Lorentz transformations (\ref{TLQPQKB}), for $ \gamma$ finite the energy of the particle would be zero in any frame; but this is a situation devoid of physical meaning as well. To preserve finite $E$ we should assign to $\gamma$ an infinite value  considering, like the previous case, $v_x=c$, the particle in motion at the speed of light. However, in order to maintain finite the wave vector $k_x$, in such a circumstance it would be also necessary 
$\omega_0 = 0$, in contradiction  with the conditions of \textit{Case 3}.

\vskip.6cm

\noindent
{\it Case 4: ${m_0=0}$\,,  \qquad ${\omega_0=0}$.}
\vskip.2cm

\noindent
Let us consider Lorentz transformations between two  inertial frames $S$ and $S'$ for the components 
of the four-momentum and the wave four-vector:

\begin{equation} \label{TLQPQKB0}
\left\{
\begin{array}{rl}
E/c &= \gamma ( E'/c + \beta p'_{x}) \\
\omega/c &= \gamma (\omega' / c  + \beta k'_{x})                                              
\end{array}
\right. \qquad 
\left\{
\begin{array}{rl}
p_{x} &= \gamma (p'_{x} + \beta E'/c)  \\
k_{x} &= \gamma (k'_{x} + \beta \omega' / c) \,.
\end{array}
\right.
\end{equation}

\noindent
If we put $m_0=0$ and $\omega_0=0$ into the invariants (\ref{invmcom}), and we consider 
in accordance with the Lemma, that $p_x$ and $k_x$ must have the same sign, we have \cite{AshMilnote}:

\begin{equation} \label{pkp}
p'_x= \frac{E'}{c}  \,, \qquad  \qquad  k_x'= \frac{\omega'}{c}  \,.
\end{equation}

\noindent
Inserting (\ref{pkp}) into (\ref{TLQPQKB0}):

\begin{equation} \label{TLQPQKBB}
\left\{
\begin{array}{rl}
E/c &= \gamma ( 1 + \beta ) E'/c \\
\omega/c &= \gamma (1 + \beta ) \omega' / c                                             
\end{array}
\right. \qquad 
\left\{
\begin{array}{rl}
p_{x} &= \gamma (1 + \beta ) E'/c  \\
k_{x} &= \gamma ( 1+ \beta) \omega' / c \,,
\end{array}
\right.
\end{equation}

\noindent
we obtain, dividing side by side:

\begin{equation} \label{rapeo}
\frac{E}{\omega}= \frac{E'}{\omega'}  \,, \qquad  \qquad  \frac{p_x}{k_x}= \frac{E'}{\omega'}   \,.
\end{equation}

\noindent
Being primed and non-primed arbitrary inertial frames, from the first of (\ref{rapeo}), we deduce that the ratio $E'/ \omega'$ has to be an invariant. As above, we define the invariant

\begin{equation} \label{Efinv}
\frac{E'}{\omega'} \equiv C \,,
\end{equation}

\noindent
and following the reasoning of \textit{Case  1}, we get again from (\ref{rapeo}) the Planck-Einstein and the de Broglie equations (\ref{relPDBL4a}). 

\vskip.6cm

\noindent
Summing up, the physically meaningful cases are \textit{Case  1} and  \textit{Case  4}. 
\textit{Case  1} corresponds to particles with inertial mass and waves with $ \omega_0\neq0$. This is the situation which is verified for fields with massive quanta. \textit{Case  4} corresponds  instead to particles without inertial mass like photons, where the associated waves have  
$\omega_0=0$, like the electromagnetic waves.


\section{Conclusions}
We have shown that, once we assume the existence of a wave phenomenon with finite energy and momentum as a classical particle, and that the momentum fluxes along the same direction of wave propagation, from Special Relativity follows that the four-momentum $p^\mu$ and the wave four-vector $k^\mu$ can be related just in one way, by a rule  of direct proportionality: 

\[ 
(E/c, {\bf p})= C (\omega/c, {\bf k}) \,,
\]
                                                                      
\noindent
where $C$ is an invariant  under Lorentz transformations.  

Therefore, we have deduced de Broglie and Planck-Einstein relations for plane waves from  more general assumptions than those used by  other autors. De Broglie and the others use the following hypotheses:

\vskip.4cm

\noindent
$1)$ The rest energy of a particle $m_0 c^2$ is proportional to a wave frequency $\nu_0$: 

\[ 
m_0 c^2=h \nu_0  \,,
\]

\noindent
where $h$ is an invariant constant.

\vskip.4cm

\noindent
$2)$ The relationship between phase velocity $v_p $ and particle velocity $v_x $ is

\[ 
v_p= \frac{v_x}{c^2}  \,,
\]

\noindent
or

\vskip.4cm

\noindent
$2')$ The frame in which the particle is at rest is the same frame in which the wave vector $k_{x}=0$. 

\vskip.4cm

\noindent
Assumptions $2)$ and $2 ')$  are equivalent and de Broglie himself shows it in his PhD thesis.
We have shown that, in order to deduce the proportionality between four-momentum
and wave four-vector, assumption  $1)$  is not a necessary starting point. 
Moreover, $2)$  and $2')$ can be replaced by another  assumption:

\vskip.5cm

\noindent
$2''')$ The momentum always flows in the same direction of the wave propagation.
\vskip.5cm

\noindent
The fact that  the particle is at rest in frame where $k_x =0$ is a situation rather arbitrary, which de Broglie justifies with his phase harmony  theorem.
But as we have shown in our work, it follows from Relativity itself, when we make the quite natural and classical assumption that wave and momentum  travel in same direction. This can be deduced by the Postulate 1.b which connects the particle with a travelling wave, if we consider the plane wave as limit case of a wave train with a broad finite length.



\begin{thebibliography}{99}
\bibitem{AshMilnote}
Of course, we could also choose
\[
p'_x=-\frac{E'}{c}   \,, \qquad   k'_x=-\frac{\omega'}{c} \,.
\]
\noindent
For brevity, we study only one of the two possibilities, the final result being the same for both of them.
\end{thebibliography}
\end{document}